\documentclass[10pt, a4paper, oneside]{article}
\usepackage{cite}
\usepackage{graphicx}
\usepackage{epsfig}
\usepackage{epstopdf}
\usepackage{amssymb,amsmath}
\usepackage[a4paper=true,pagebackref=false]{hyperref}
\usepackage[scale=0.75]{geometry}
\usepackage[labelfont=bf,textfont=md]{caption}
\hypersetup{colorlinks = true, linkcolor = red, anchorcolor = red, citecolor = cyan, filecolor = magenta, pagecolor = red, urlcolor = red}

\title{Cosmic acceleration and ekpyrotic bounce with Chameleon field}

\author{Ashutosh Singh$^{1}$\footnote{ashuverse@gmail.com}, Anirudh Pradhan$^{1}$\footnote{pradhan.anirudh@gmail.com}
\\ \\
${}^{1}$Centre for Cosmology, Astrophysics and Space Science, \\ GLA University, Mathura, Uttar Pradesh 281406, India}

\date{}
\begin{document}
\maketitle

\begin{abstract}
In this article, we explore the homogeneous and isotropic flat Friedmann-Robertson-Walker (FRW) model in Chameleon cosmology. By considering a non-minimal coupling between the scalar field and matter, we present a non-singular bouncing cosmological scenario of the universe. The universe initially exhibits the ekpyrotic phase during the contracting era, undergoes a non-singular bounce, and then in expanding era, it smoothly transits to the decelerating era having matter and radiation dominated phases. Further, this decelerating era is smoothly connected to the 
late-time dark energy-dominated era of the present epoch. We use numerical solution techniques to solve non-minimally coupled gravity equations 
for understanding the evolution of scalar field along with other quantities like effective potential in the model. The model thus unifies an 
ekpyrotic, non-singular, asymmetric bounce with the dark energy era of the present epoch. We study the evolution of bouncing model and confront 
the model with observational results on the equation of state parameter by constraining the model parameters.
\end{abstract}



\noindent\hrulefill
\section{Introduction}
In Einstein's General relativity (GR), late-time cosmic acceleration may be explained by an additional degree of freedom consisting of 
barotropic fluid/scalar field with negative pressure termed as dark energy. The dark energy-domination of the present epoch has been the driving factor behind the late-time accelerated expansion of universe, which is confirmed from the observational evidences of different 
probes \cite{Riess+1998,Perl+1999,Ade+2014}. A positive cosmological constant serves as the simplest candidate for dark energy in GR, but 
it suffers many problems. So, in literature, other candidates for dark energy have also been proposed; see reviews \cite{Capozziello+2019,Bamba+2012}. 
The models consist of cosmological constant, modified gravity theories like $f(R)$ gravity, and barotropic fluids having the form $p=F(\rho)$ 
yields the negative pressure component and thus explains the late-time accelerating behavior of the universe 
\cite{Nojiri+2005,Nojiri+72+2005,Quercellini+2007,picon2000,chiba2000,fas2010,noo2017}. However, many of these models have problems of their own, 
which can be found in reviews \cite{Capozziello+2019,Bamba+2012,noo2017}.\\ 
Among various proposals for the explanation of dark energy, Khoury and Weltman \cite{kw2004a,kw2004b} have suggested the Chameleon cosmology where
a scalar field couples non-minimally to matter through a conformal factor with gravitational strength and thus, its mass depends sensitively
on the environmental situations. That is, the mass of the scalar field is varying and depends on the local matter density. In regions of high density,
the chameleon field intermixes with its environment and becomes invisible for the equivalence principle searches \cite{bkw2004}. It becomes very
light in the diluted matter situations and thus this scalar field may also explain the late-time accelerating universe by playing the role of
dark energy \cite{kw2004a,kw2004b,bkw2004}. Chameleon scalar field is also consistent with the constraints on existence of non-minimally coupled
scalars \cite{kw2004b,bkw2004}. The chameleon mechanism may overcome quintessence mechanism drawbacks and act as a dark energy candidate
for late-time cosmic acceleration. The large-scale behavior of the chameleon field has been investigated in literature \cite{1r,1s}. These different
behaviors on large and small scales depend on the potential function and coupling function. In chameleon field cosmology, these functions
contribute to the effective mass of the scalar field \cite{kw2004a,kw2004b}. The different theoretical and observational aspects including
thermodynamic and inflationary behaviors have been discussed in detail for chameleon scalar field cosmology
\cite{1r,1s,1l,1m,1o,1p,1q,1t,1u,1v,1vx,1x,1y,1z}.\\
The question regarding the universe starting its journey by expanding from the big bang singularity or from a stage free from singularity
(like the non-singular bounce) is an interesting one. The inflationary era in the standard cosmological model signifies from exponential evolution scenarios like de Sitter or quasi-de Sitter. Odintsov and Oikonomou have explored a pre-inflationary non-singular bounce exhibiting a subsequent quasi-de Sitter era \cite{revi5}. The pre-inflationary bounce may avoid the past cosmic singularity. Our aim is to study the evolution of ekpyrotic scenario and non-singular bounce
for the chameleon scalar field, yielding a late-time accelerated universe. In other words, we aim to model the universe with the non-singular bounce
as the initial stage of the expanding era. Further, we use the numerical technique approach to investigate other aspects of the evolving universe.
We show that the scale factor of the model unifies the ekpyrotic contracting universe with the late-time dark energy-dominated era of the present universe.
We write the cosmological equations in section \ref{sec2}. In section \ref{sec3}, we study the different issues like the background evolution of the
universe in the model along with scrutiny of the theoretical effective equation of state parameter with observational findings. In section \ref{conclusion},
we provide the summary of obtained results.

\section{Chameleon model equations}
\label{sec2}
The metric for homogeneous and isotropic flat universe is given by
\begin{equation}
	ds^{2}=dt^{2}-a^{2}(t)\left(dr^{2}+r^{2}\left(d\theta^{2}+\sin^{2}\theta d\phi^{2}\right)\right)
	\label{eq1}
\end{equation}
where $a(t)$ is the scale factor. We take the chameleonic gravity action, where matter sector is non-minimally coupled with scalar field, such 
that the action of model is given by \cite{fas2010}
\begin{equation}
	S=\int \left(\frac{R}{16\pi G}+\frac{1}{2}\phi_{,i}\phi^{,i}-V(\phi)+f(\phi)L_m \right)\sqrt{-g}d^4x
	\label{eq2} 
\end{equation}
where $L_m$ is the Lagrangian of the standard matter field, $R$ is the Ricci scalar constructed from metric, $V(\phi)$ and $f(\phi)$ are potential
and some analytic function for chameleon scalar field $\phi$. Varying the action (\ref{eq2}) with respect to the metric component yield the field equations for flat FRW metric as
\begin{equation}
	3H^{2}=\rho f+\frac{1}{2}\dot{\phi}^2+V(\phi), \quad 2\dot{H}+3H^{2}=-p f-\frac{1}{2}\dot{\phi}^2+V(\phi)
	\label{eq3}
\end{equation}
where, we take $8\pi G=c=\hbar=k_b=1$. Here, $\rho$ stands for energy density contribution for matter and we take $p=\gamma\rho$. 
Hubble parameter is defined by $H=\frac{\dot{a}}{a}$. Overhead dot and prime denotes derivatives with respect to $t$ and $\phi$ respectively. 
Varying the action (\ref{eq2}) with respect to scalar field $\phi$ yields the wave equation for the chameleon field as
\begin{equation}
	\ddot{\phi}+3H\dot{\phi}+V'+\frac{1}{4}(\rho-3p)f'=0
	\label{eq4}
\end{equation}
From equations (\ref{eq3}) and (\ref{eq4}), the continuity equation for matter takes the form
\begin{equation}
	\dot{\rho}+3H(\rho+p)+\frac{3}{4}\frac{\dot{f}}{f}(\rho+p)=0
	\label{eq5}
\end{equation}
Solving above equation (\ref{eq5}), we may write 
$\rho=\left(\frac{a_0}{a} \right)^{3(1+\gamma)}\left(\frac{f_0}{f(\phi)} \right)^{\frac{3}{4}(1+\gamma)}$, where $a_0$ and $f_0$ are constant parameters.\\
In comparison to standard Friedmann equations of GR, we may write 
$\rho_e= \rho f+\frac{1}{2}\dot{\phi}^2+V(\phi)$ and $p_e=p f+\frac{1}{2}\dot{\phi}^2-V(\phi)$ and define effective equation of state (EoS) 
parameter as $\gamma_e=\frac{p_e}{\rho_e}$. 

\section{Background evolution and dynamical properties of the non-singular ekpyrotic bouncing model}
\label{sec3}
\begin{figure}[h!]
	\centerline{\includegraphics[width=2.25in]{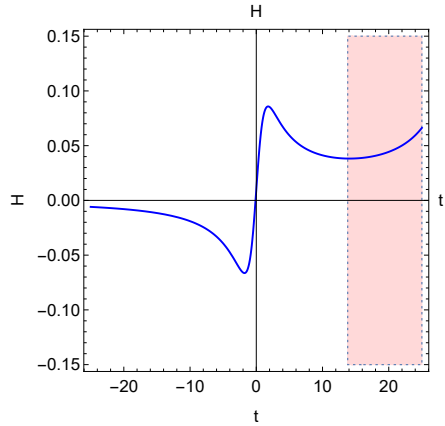}}
	\vspace*{8pt}
	\caption{$H$ with $t$, $t$ is given in Billion years\protect\label{fig1}}
\end{figure}

\begin{figure}[h!]
	\centerline{\includegraphics[width=2.25in]{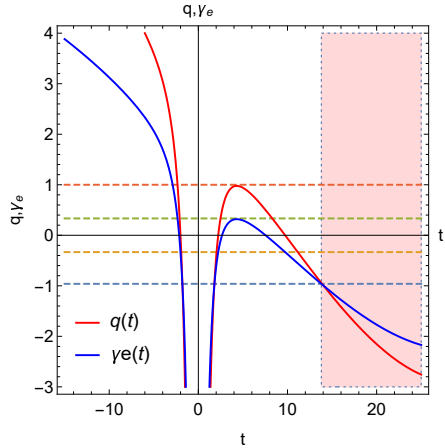}}
	\vspace*{8pt}
	\caption{$q$ and $\gamma_e$ with $t$, $t$ is given in Billion years and dashed lines representing $y=1$, $y=\frac{1}{3}$, $y=-\frac{1}{3}$ 
		and $y=-0.961$ respectively \protect\label{fig2}}
\end{figure}

\begin{figure}[h!]
	\centerline{\includegraphics[width=2.25in]{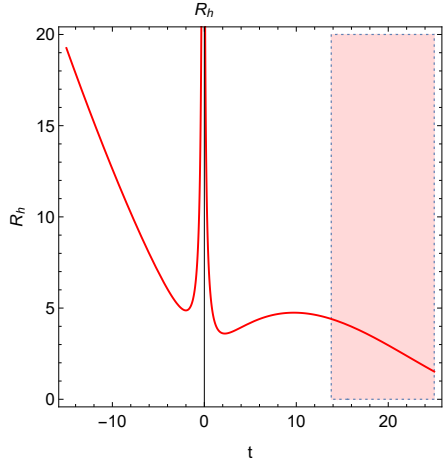}}
	\vspace*{8pt}
	\caption{$R_h=\mid aH\mid^{-1}$ with $t$, $t$ is given in Billion years\protect\label{figrh}}
\end{figure}

In this section, we investigate the dynamical features of the chameleon field by considering a certain form of the scale factor. In particular,
we study two different eras, namely near the non-singular bounce point and at the late-time epoch. The bouncing universe evolution consists
of an era of contraction followed by an expansion era with minima in scale factor at bounce instant. The bounce in FRW model is characterized
by $\dot{a}(t_b)=0$ with $\ddot{a}>0$ in small neighborhood of $t=t_b$, where $t_b$ denotes the bounce instant \cite{noo2017,1ae,1ab,1af,1ag,1ah,1ac}.
For non-singular bounce, we should have $a(t_b)\neq 0$. To unify an ekpyrotic non-singular bounce with dark energy dominated era, we
take the scale factor as \cite{8a,8b}
\begin{equation}
	a(t)=\left[ 1+a_0\left(\frac{t}{t_0} \right)^2 \right]^n\exp\left( \frac{1}{\beta-1}\left(\frac{t_s-t}{t_0} \right)^{1-\beta} \right)
	\label{eq6}  
\end{equation}
where $n,\beta,t_s$ are constants and $t_0$ is the fiducial time, taken for dimensional correctness. We take $t_0=1$ Billion years (By) 
in the present calculation. We may write equation (\ref{eq6}) as
\begin{equation}
	a(t)=\left[ 1+a_0t^2 \right]^n\exp\left( \frac{1}{\beta-1}(t_s-t)^{1-\beta} \right)
	\label{eq7}  
\end{equation}
With $a_1(t)=\left[ 1+a_0t^2 \right]^n$ for $n<\frac{1}{6}$, we may get a non-singular ekpyrotic bounce scenario. For large positive time 
$t$, $a_1(t)$ will behave like $t^{2n}$, which is inconsistent with a viable late-time dark energy era.\\
Ekpyrotic bounce may be unified with a viable dark energy era at late-times by taking $a(t)=a_1(t)\times a_2(t)$ with $n<\frac{1}{6}$, 
where $a_2(t)=\exp\left( \frac{1}{\beta-1}(t_s-t)^{1-\beta} \right)$. The $a_2(t)$ expression has almost no role in the contracting era and 
the bouncing behavior will be controlled by $a_1(t)$. However, the presence of $a_2(t)$ will modify the bouncing instant of the universe. 
$a_2(t)$ will be important during the expanding phase of the universe. The $a_2(t)$ term along with $a_1(t)$ will lead to a viable current 
dark energy era/late-time accelerating era of the universe. The scale factor $a(t)$ appears to unify the dark energy era from an ekpyrotic 
bounce with an intermediate decelerating phase in-between bounce and late-time acceleration era, which may be confirmed by studying the behavior 
of effective EoS parameter $\gamma_e$.\\
The Hubble parameter measures the expansion rate of the universe and may be written for the present scale factor (\ref{eq7}) as
\begin{equation}
	H(t)=\frac{2a_0nt}{1+a_0t^2}+\frac{1}{(t_s-t)^\beta}
	\label{eq8}
\end{equation}
From above expression, we may observe that the Hubble parameter and/or it's higher derivatives diverges at $t=t_s$, depending on the value of 
$\beta$. Finite time future singularities \cite{Nojiri+2005,Nojiri+72+2005} may be characterized on the basis of $\rho_e$ and $p_e$ 
(equivalently, in terms of Hubble parameter and it's derivatives) in the cosmological model. Type-I singularity is present in model for 
$\beta>1$. For $0<\beta<1$, the model at $t=t_s$ will undergo with Type-III singularity. For $-1<\beta<0$, Type-II singularity appear at 
$t=t_s$. For $\beta<-1$ and non-integer, a Type-IV singularity is present in the model. Type-IV singularity \cite{revi1,revi4,revi2,revi3} may also be termed as `soft' type of singularity, since the universe can smoothly pass through this singularity. The physical quantities like energy density and pressure also remain finite at this singularity. However, the Hubble parameter derivatives $H^{(n)}(t),n\geq 2$ may diverge at this singularity. Inflationary \cite{revi4} as well as bouncing models \cite{revi1,revi2} have been explored in literature having Type-IV singularity. The Type-IV singularity may cause dynamical instabilities. The occurrence of Type-IV singularity at the bounce point may lead to a singular bounce \cite{revi1}. In the present paper, in order to unify an ekpyrotic contracting scenario with the late-time expanding era, we take the values of model parameters in such a way that the model will have Type-I singularity in future. We may conclude that a finite-time 
singularity will be present in the model, irrespective of the value of $\beta$. Up to the present epoch, $t_p\approx 13.8$ By, we may 
describe a singularity-free evolution by taking $t_s>t_p$. This consideration of $t_s>t_p$ provides a singularity free evolution - at 
least up to - the present epoch of the universe. \\
For positive $a_0,n,\beta$, the Hubble parameter $H(t)>0$ for $t>0$. For $t<0$, we may observe that $\frac{2a_0nt}{1+a_0t^2}$ term of $H(t)$ 
will be negative while $\frac{1}{(t_s-t)^\beta}$ term will be always positive. Therefore, the condition of bounce ($H(t_b)=0,\dot{H}>0$) may 
be satisfied for some negative $t$. For $t<0$, we may write $t=-\mid t\mid$ and equation (\ref{eq8}) may be written as
\begin{equation}
	H(t)=-H_1(t)+H_2(t)
	\label{eq9}
\end{equation}
with $H_1(t)=\frac{2a_0n\mid t\mid }{1+a_0\mid t\mid^2}$ and $H_2(t)=\frac{1}{(t_s+\mid t\mid)^\beta}$. We may observe that for 
$t\rightarrow -\infty$, both $H_1(t),H_2(t)$ starts from zero value and are increasing function of time. $H_1(t)$ increases at faster rate 
as compared to $H_2(t)$. At $t=0$, we have $H_1(t)=0$, but $H_2(t)$ is positive. It simply means that there is a negative $t$, for example 
$t=-\tau$ with $\tau>0$ where $H(t=-\tau)=0$. Following conclusions may be drawn on the basis of behavior of $H(t)$ at $t=-\tau$:
\begin{enumerate}
	\item $H_1(t)>H_2(t)$, that is, the universe is contracting during $t<-\tau$.
	\item $H_1(t)=H_2(t)$, that is, the bounce instant at  $t=-\tau$.
	\item $H_1(t)<H_2(t)$, that is, the universe is expanding during $t>-\tau$.
\end{enumerate}
The algebraic equation $H(-\tau)=0$ may not be solved in closed form for $\tau$. For $a_0=0.32,n=\frac{2}{15},\beta=1.315,t_s=34$ 
the expression $\frac{2a_0n\tau}{1+a_0\tau^2}=\frac{1}{(t_s+\tau)^\beta}$ will yield bounce instant $t=-\tau=-0.113467$. And, $a(t_b)=2.84292$ 
which is positive value, thus indicating the non-singular bounce. Background evolution of late-time contracting era, the scale factor and 
Hubble parameter behaves as $a(t)\approx {a_0}^nt^{2n}$ and $H(t)\approx \frac{2n}{t}$. The effective EoS takes the form 
$\gamma_e=-1+\frac{1}{3n}=\gamma$. Therefore, for ekpyrotic nature of bounce, $\gamma>1$ (equivalently $n<\frac{1}{6}$). With this constraint 
on $n$, the chameleon field in the contraction era decays faster than the anisotropic energy density. The background evolution during the 
contracting era remains stable against the growth of anisotropies, therefore the model is free from the BKL instability. The behaviors of 
$H(t)$ and effective EoS parameter along with deceleration parameter have been given in figure (\ref{fig1}) and (\ref{fig2}) respectively.\\
From figure (\ref{fig1}), it may be realized that the universe undergoes bounce in negative $t$ zone, in particular at $t=-0.113467$ which 
lies in interval $t\in(-0.114,-0.113)$. In positive time zone, one may observe that the Hubble parameter becomes an increasing function of 
time in region $t>t_p$, which will plague the the present model with future singularity.\\
Figure (\ref{fig2}) highlights the behavior of the deceleration parameter and effective EoS parameter. The deceleration parameter describes the rate at
which the universe expansion is slowing down \cite{1aa,2022a}. Using the geometrical sector in Equation (\ref{eq3}), we may equivalently define the effective
EoS parameter as $\gamma_e=-1-\frac{2\dot{H}}{3H^2}$. In the contracting era, especially away from bounce instant, $\gamma_e>1$ and thus highlighting
the ekpyrotic behavior of the present model. Near the bouncing instant in expanding era, after the bounce, the model yields a phantom 
energy-dominated era which smoothly decays into a radiation-dominated era, further decaying into the matter-dominated era. The decelerating universe 
crosses $q=0$ line, and with this crossing, the universe starts accelerating; thus, a dark energy-dominated era may be observed in the model. Therefore,
a late-time accelerating universe may be realized in the present non-singular bouncing model with the decelerating past of matter and radiation-dominated
eras. The model thus realizes a complete cosmological scenario (initial accelerated expansion may be called inflation followed by
radiation and matter-dominated eras further decaying into dark energy-dominated eras at late times) during its evolution.\\
Deceleration parameter $(q)$ may be defined as $q=-1-\frac{\dot{H}}{H^2}$ and takes the form in present model as
\begin{equation}
	q=-1-\frac{(t_s-t)^{\beta-1}\left(\beta(1+a_0t^2)^2-2a_0n(a_0t^2-1)(t_s-t)^{\beta+1} \right) }{\left[ a_0t(2n(t_s-t)^{\beta}+t)+1\right]^2 }
	\label{eq10}
\end{equation}
The de Sitter expansion of the universe may be realized at $q=-1$, and for $q<-1$, the universe undergoes super-exponential expansion. 
The matter and radiation lines are $q=\frac{1}{2}$ and $q=1$ respectively. For $-1<q<0$, the universe undergoes accelerating power-law expansion. 
The above expression of $q(t)$ may be used to develop an understanding of the accelerating or decelerating stage of the universe. 
The transition from deceleration to acceleration stage (and vice versa) may be described by taking $\ddot{a}=0$, which in turn means that $q=0$. 
By using the values of model parameters, we may identify that $q=0$ at $t=t_1=2.22852$ and $t=t_2=9.73692$ in the expanding era of universe. 
At $t=t_1$, the model enters into decelerating stage from initial era of expansion and at $t=t_2$, the model transits from decelerating era 
into accelerating era, realizing current dark energy dominated era. The second transition from $q>0$ to $q<0$ may be visualized as late-time 
accelerating era of the present universe. For selected values of model parameters, it may be identified that $t_2<t_p$, where $t_p$ denotes 
the present age of universe. For $n<\frac{1}{6}$ or equivalently $\gamma>1$ the present model suffers from Type-I singularity at $t=t_s$. 
Therefore, for singularity-free universe evolution up to present cosmic time $t_p$, the parameter $t_s$ needs to satisfy the constraint $t_s>t_p$. 
An upper bound on $t_s$ may be introduced by qualitative estimation of $t_2$ from definition of $q(t)$ and is given by $t_p<t_s<2\beta t_p$ \cite{8b}.\\
We confront the theoretical EoS of present model with Planck+SNe+BAO result. The effective EoS parameter at present epoch has been 
constrained to be $\gamma_e(t_p)=-0.957\pm 0.080$ \cite{netal}. For the specified values of model parameters, the effective EoS of the present 
model is $\gamma_e(t_p)=-0.961$, which is compatible with the observational results. The pictorial depiction of $\gamma_e$ can be seen from 
figure (\ref{fig2}).\\
It is interesting to note that in the present model, the ekpyrotic phase during contraction resolves the anisotropy problem of the bouncing
cosmologies. This anisotropy problem is also known as BKL instability. During the contracting phase, the anisotropic energy density of the background
metric is used to grow faster than the energy density of the bounce-inducing component, leading to a scenario where background evolution becomes
unstable, which has been termed BKL instability \cite{bkl}.\\
The behavior of cosmological perturbations generated in the present model may be revealed by evolution of co-moving Hubble radius $R_h=\frac{1}{\mid aH\mid}$. The behavior of co-moving Hubble radius with time has been shown in figure (\ref{figrh}). Due to the vanishing Hubble parameter at bounce, $R_h$ diverges. In the contracting era, the co-moving Hubble radius diverges as $t\rightarrow -\infty$. With contraction of universe, $R_h$ also decreases and have a minimum before diverging at the bounce instant. It means that the primordial perturbation modes generated in the deep contracting era will lie within the deep sub-Hubble region. In the expanding era after bounce, $R_h$ monotonically decreases with increasing time, pointing towards a late-time accelerated universe era.  
\subsection{Dynamic evolution of scalar field and effective potential}
\begin{figure}[h!]
	\centerline{\includegraphics[width=2.25in]{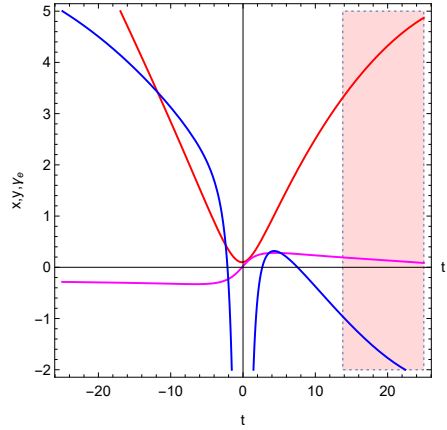}}
	\vspace*{8pt}
	\caption{$\phi$ (red curve), $\dot{\phi}$ (magenta curve) and $\gamma_e$ (blue curve) with $t$, $t$ is given in Billion years\protect\label{fig3}}
\end{figure}

\begin{figure}[h!]
	\centerline{\includegraphics[width=2.25in]{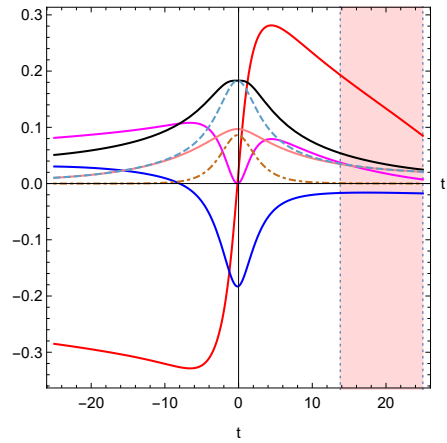}}
	\vspace*{8pt}
	\caption{$\dot{\phi}$ (red), $\dot{\phi}^2$ (magenta), $\rho_e$ (black), $p_e$ (blue), $\rho f$ (pink), $V$ (dot-dashed) and $V_e$ (dashed) 
		with $t$, $t$ is given in Billion years\protect\label{fig4}}
\end{figure}

In order to solve the scalar field equation, we take the potential to be of form $V=V_0e^{\mu\phi}$ and assume that the function $f(\phi)$ behaves 
exponentially as $f(\phi)=f_0e^{\lambda\phi}$, where $V_0,f_0,\mu,\lambda$ are constant parameters. These choices are purely phenomenological, 
which are leading to different behavior of universe including phantom crossing scenario of the model. In the present discussion, we adopt the 
numerical techniques to solve equation (\ref{eq4}) for the above choices of potential and non-minimal coupling function. By writing 
$\phi=x,\dot{\phi}=y$, we solve equation (\ref{eq4}) for $f_0=0.1,\lambda=-0.32,V_0=0.1,\mu=-1.5$ with $a_0=0.32,\beta=1.315,n=\frac{2}{15},t_s=34$ 
and initial conditions $x(t_b)=0.1,y(t_b)=0.05$, where $t_b=-0.1134$. The observational evidences suggest that the present day universe consists 
mostly of components having EoS $\gamma=0,\gamma\approx -1$ and radiation is negligible as compared to these matter and dark energy dominated 
components \cite{netal}. So, for solving the equations, we take the universe composed of matter and dark energy dominated components. The numerical 
solution highlighting the behavior of $\phi,\dot{\phi}$ along with $\gamma_e$ has been given in figure (\ref{fig3}), with shaded region highlighting 
the time $t>t_p=13.8$ By. The scalar field achieves minima of its value near the bounce instant and shows bouncing behavior. This behavior 
is also supported by behavior of $\dot{\phi}$. The bounce in scalar field is sensitive to the the choice of initial conditions. 
With increasing (decreasing) $a(t)$, we may see the increasing (decreasing) behavior of scalar field $\phi$.\\
By using the scalar field numerical solution, we plot the behaviors of $\dot{\phi},\dot{\phi}^2,\rho_e,p_e,\rho f,V, V_e$ in figure (\ref{fig4}), 
where $V_e=V+\rho f$, $\rho_e=\rho f+\frac{1}{2}\dot{\phi}^2+V$ and $p_e=\rho f+\frac{1}{2}\dot{\phi}^2-V$. It may be observed from 
figure (\ref{fig4}) that the quantities $\rho_e,\rho f,V,V_e$ attains their maxima whereas $p_e,\dot{\phi}^2$ attains their minima at the bounce.
\begin{figure}[h!]
	\centerline{\includegraphics[width=2.25in]{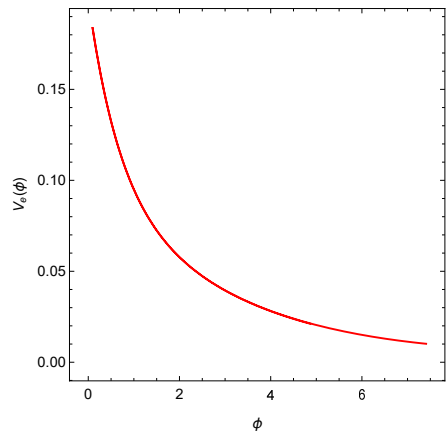}}
	\vspace*{8pt}
	\caption{Effective potential $V_e(\phi)$ with field $\phi$ \protect\label{fig4a}}
\end{figure}
One may find the mass associated with the field $\phi$ from expression \cite{1o}
\begin{equation}
	m^2=\frac{d^2}{d\phi^2}V_e(\phi)
	\label{eq11}
\end{equation}
The effective potential monotonically increases during the contracting phase and monotonically decreases to a minimum value with the evolution of 
time during the expanding phase. On the other hand, the field also shows bouncing behavior, as it is monotonically decreasing during the 
contracting phase, with minima near the bounce, and starts monotonically increasing in expanding phase. The behavior of effective potential 
$V_e(\phi)$ with $\phi$ has been given in figure (\ref{fig4a}). With increasing $\phi$, $V_e(\phi)$ tends to a minimum value.
\subsection{Role of energy conditions}
\begin{figure}[h!]
	\centerline{\includegraphics[width=2.25in]{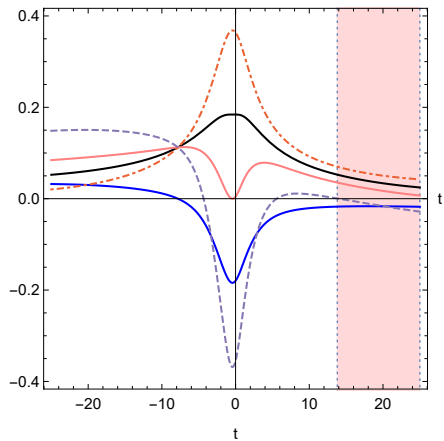}}
	\vspace*{8pt}
	\caption{$\rho_e$ (black), $p_e$ (blue), $\rho_e+p_e$ (pink), $\rho_e-p_e$ (dot-dashed) and $\rho_e+3p_e$ (dashed) with $t$, $t$ is 
		given in Billion years\protect\label{fig5}}
\end{figure}

Energy conditions play an interesting role during the cosmological evolution of the universe. The universe's evolution, in particular, accelerating
or decelerating behavior and the emergence of future singularities can be linked to the constraints imposed by energy conditions.
By considering the formulation of point-wise energy conditions at a point in space-time in GR, we may formulate the expression of null
energy condition (NEC), Weak energy condition (WEC), Dominant energy condition (DEC) and Strong energy condition (SEC) in the effective
formulation of the present model. NEC is given by $\rho_e+p_e\geq 0$. WEC is given by $\rho_e\geq 0,\rho_e+p_e\geq 0$. DEC is given by
$\rho_e\geq 0,\rho_e\pm p_e\geq 0$. SEC is given by $\rho_e+p_e\geq 0,\rho_e+3p_e\geq 0$. Note that the violation of NEC will violate all
other energy conditions. We highlight the behavior of these energy conditions along with behavior of $p_e$ in figure (\ref{fig5}), with shaded
region highlighting the era for $t>t_p=13.8$ By. Observe that during ekpyrotic era, WEC, NEC and DEC are satisfied except $\rho_e+3p_e\geq 0$.
Near the bounce instant, for a very brief period of time, $\rho_e+p_e<0$, that is, NEC is not satisfied near the bounce. However, during expanding era,
all energy conditions are satisfied except $\rho_e+3p_e\geq 0$. From Raychaudhuri-like equation, in effective scenario, $\rho_e+3p_e< 0$
highlights the fact that the dark energy component is dominating upon other components of an effective fluid and thus leading to a late-time
accelerating universe. The inequality $\rho_e-p_e\geq 0$ in the model suggests that the adiabatic sound speed squared ${c_s}^2<1$, leading
to a causal universe. However, $\rho_e+p_e< 0$ near the bounce will cause a classically unstable universe.

\section{Conclusions}
\label{conclusion}
In the present manuscript, we investigate the behavior of chameleon scalar field, which is non-minimally coupled with the matter. Due to the
interaction of scalar field with the matter sector, we get a generalized energy conservation law in the present model. Evolution of the scale
factor yields a non-singular bouncing model, which is asymmetric in nature. The model is composed of ekpyrotic contraction and late-time
accelerating expansion of the universe. In the expanding era, the late-time accelerating expansion of the universe has a decelerating past of
a radiation-dominated era followed by a matter-dominated era. These transitions from accelerating to decelerating and decelerating to accelerating era have
been smooth during the universe's evolution. We constrain the model parameters, especially the effective EoS parameter, with observations of
Planck+SNe+BAO \cite{netal} and found that the present value of $\gamma_e$ lies in range $-0.957\pm 0.80$. The model also predicts
finite-time future singularity of Type-I nature based on values of model parameters.\\
The numerical solutions for appropriate choices of model parameters and initial conditions have been displayed graphically for the evolution
of the scalar field along with other quantities like effective energy density, effective pressure, and effective potential. The chameleon scalar
field exhibits a bouncing nature during its evolution in the model. The effective potential tends to a minimum value during its evolution
in the late-time contracting era and late-time expanding era. The model satisfies NEC, WEC, DEC away from bounce, but in a very small
neighborhood of bounce instant, NEC is not satisfied and, thus, violates all other energy conditions at the bounce. In the
expanding era, $\rho_e+3p_e\geq 0$ is violated during the accelerated expansion of universe. From the validity of DEC, it may be concluded
that the model remains classically stable during contracting and expanding era except for a brief period of time near bounce.  \\
We may conclude that the present model can unify the dark energy-dominated era of the present universe, with the ekpyrotic contracting
era having a non-singular bounce. The model is free from BKL instability. At the same time, it is compatible with the present-day observational
value of effective EoS parameter, subjected to the constraints on model parameters. It would be interesting to investigate the chameleon
scalar field dynamics with negative potentials \cite{trs2018}, we leave it for a future study.

\section*{Acknowledgments}

We are grateful to the honorable reviewer for the valuable comments and suggestions for the modification of paper. A. Pradhan thanks the IUCAA, Pune for providing facilities and support under visiting associateship program.

\end{document}